\documentclass{PoS}

\title{Kaon Physics: Theory Overview}

\ShortTitle{Kaon Physics}

\author{\speaker{Antonio Pich}\\ 
        IFIC, Universitat de Val\`encia -- CSIC, Apt. Correus 22085, E-46071 Val\`encia, Spain \ \& \\
        Physik Dep. and Inst. for Advanced Study, Tech. Univ. M\"unchen, D-85748 Garching, Germany\\
        E-mail: \email{Antonio.Pich@ific.uv.es}}

\abstract{Kaon decays have played a key role in the construction of the Standard Model and continue to be
an important testing ground of the electroweak flavour theory. They can provide new signals
of CP-violation phenomena and, perhaps, a window into physics beyond the Standard Model.
The interplay of long-distance QCD effects in strangeness-changing transitions can be analyzed,
combining the short-distance Operator Product Expansion with Chiral Perturbation Theory techniques.
A brief overview is presented, focusing on a few selected decay modes.
A more detailed and comprehensive review can be found in Ref.~\cite{Cirigliano:2011ny}.}

\FullConference{The XIth International Conference on Heavy Quarks and Leptons,\\
		June 11-15, 2012\\
		Prague, Czech Republic}

\def\be{\begin{equation}}
\def\ee{\end{equation}}
\def\beqn{\begin{eqnarray}}
\def\eeqn{\end{eqnarray}}

\def\ba{\begin{array}{c}}
\def\bat{\begin{array}{cc}}
\def\ea{\end{array}}
\def\bi{\begin{itemize}}
\def\ei{\end{itemize}}

\def\cL{{\cal L}}

\def\cH{{\cal H}}

\def\cO{{\cal O}}

\newcommand{\eqn}[1]{(\ref{#1})}
\newcommand{\bel}[1]{\be\label{#1}}

\newcommand{\lsim}{~{}_{\textstyle\sim}^{\textstyle <}~}
\newcommand{\rms}{\rm\scriptsize}

\begin{document}

\section{Theoretical Framework}

Kaons have been at the center of many fundamental developments in particle physics,
playing a key role in the construction of what we now call the Standard Model (SM):
the introduction of internal flavour quantum numbers (strangeness) \cite{GellMann:1953zza,Pais:1952zz},
parity violation \cite{Dalitz:1954cq,Lee:1956qn},
meson-antimeson mixing \cite{Lande:1956pf,Fry:1956pg},
quark mixing \cite{Cabibbo:1963yz,Kobayashi:1973fv},
the discovery of CP violation \cite{Christenson:1964fg},
and the suppression of flavour-changing neutral currents and the GIM mechanism \cite{Glashow:1970gm}.
High-precision experiments on rare kaon decays provide sensitivity to short-distance scales ($c$, $t$, $W^\pm$, $Z$)
and offer the exciting possibility of unravelling new physics beyond the SM.
Searching for forbidden flavour-changing processes
beyond the $10^{-10}$ level
[Br$(K_L\to e^\pm \mu^\mp) < 4.7\times 10^{-12}$ \cite{Ambrose:1998us},
Br$(K_L\to e^\pm e^\pm \mu^\mp\mu^\mp) < 4.12\times 10^{-11}$ \cite{AlaviHarati:2002eh},
Br$(K^+\to\pi^+\mu^+ e^-) < 1.3\times 10^{-11}$ \cite{Sher:2005sp},
Br$(K^+\to\pi^+\mu^- e^+) < 5.2\times 10^{-10}$ \cite{Appel:2000tc} (90\% C.L.)],
one is actually exploring energy scales above the 10 TeV region.
The study of allowed (but highly suppressed) decay modes
provides, at the same time, very interesting tests of the SM
itself. Electromagnetic-induced non-leptonic weak transitions
and higher-order weak processes are a useful tool to improve our
understanding of the interplay among electromagnetic, weak and strong
interactions. In addition, new signals of CP violation, which would
help to elucidate the source of CP-violating phenomena, can be looked for.

The theoretical analysis of kaon decays is highly non-trivial. While the
underlying flavour-changing transitions among the constituent
quarks are associated with the electroweak scale, the
corresponding hadronic amplitudes are governed by the long-distance
behaviour of the strong interactions, {\it i.e.}, the confinement regime of QCD.
The short-distance approach to weak transitions makes
use of the asymptotic freedom property of QCD
to successively integrate out the fields with heavy masses down to
scales $\mu < m_c$.
Using the operator product expansion (OPE) and
renormalization-group techniques, one gets an effective $\Delta S=1$ Hamiltonian \cite{Gilman:1979bc}
\bel{eq:sd_hamiltonian}
\cH_{\mbox{\rms eff}}^{\Delta S=1} \; = \; {G_F\over\sqrt{2}}\,
V_{ud}^{\phantom{*}} V_{us}^*\;
\sum_i\, C_i(\mu)\, Q_i \, + \, \mbox{\rm h.c.},
\qquad\qquad\quad
C_i(\mu)\; =\; z_i(\mu) -  y_i(\mu)\, \frac{V_{td}^{\phantom{*}}V_{ts}^*}{V_{ud}^{\phantom{*}} V_{us}^*}\, ,
\ee
which is a sum of local four-fermion operators $Q_i$,
constructed with the light degrees of freedom
($u,d,s; e,\mu,\nu_l$), modulated by
Wilson coefficients $C_i(\mu)$
which are functions of the heavy ($Z,W,t,b,c,\tau$) masses.
The CP-violating decay amplitudes are proportional to the components $y_i(\mu)$.
The overall renormalization scale $\mu$
separates the short- ($M>\mu$) and long-distance ($m<\mu$)
contributions,  which are contained in $C_i(\mu)$
and $Q_i$, respectively.
The Wilson coefficients are fully known at the next-to-leading order (NLO)
\cite{Buras:1992tc,Ciuchini:1995cd};
this includes all corrections of $\cO(\alpha_s^n t^n)$ and $\cO(\alpha_s^{n+1} t^n)$,
where $t\equiv\log{(M_1/M_2)}$ refers to the logarithm of any ratio of
heavy mass scales ($M_{1,2}\geq\mu$).
In order to calculate the kaon decay amplitudes,
we also need to know the nonperturbative matrix elements of the operators
$Q_i$ between the initial and final states.

The low-energy strong interactions are better understood with symmetry considerations, because
the lightest pseudoscalar mesons correspond to the octet of Goldstone bosons associated
with the dynamical chiral symmetry breaking of QCD: $SU(3)_L\otimes SU(3)_R \rightarrow SU(3)_V$.
Their dynamical properties can then be worked out systematically
through an effective Lagrangian.
The quark and gluon fields of QCD are replaced by a unitary
matrix \ $U(\phi) \equiv \exp(i \sqrt{2} \Phi /F)$,
parameterizing the Goldstone excitacions over the vacuum quark
condensate  $\langle\bar{q}_L^j q_R^i\rangle$ ($i,j=u,d,s$).
The Chiral Perturbation Theory  \cite{Weinberg:1978kz,Gasser:1984gg,Ecker:1994gg,Pich:1995bw} ($\chi$PT)
formulation of the Standard Model is given by the most general effective Lagrangian, involving the matrix
$U(\phi)$, which is consistent with chiral symmetry.
The Lagrangian can be organized in terms of increasing powers of momenta (derivatives)
and quark masses over the chiral symmetry-breaking scale ($\Lambda_\chi\sim 1$~GeV).
All short-distance information is encoded in the low-energy couplings (LECs) of the $\chi$PT operators.
At lowest order (LO), $\cO(p^2)$, the strong interactions are fully parameterized in terms of only two LECs:
the pion decay constant $F \simeq F_{\pi}=92.4$ MeV and $B_0\simeq -<\bar{u} u>/F^2$,
which accounts for the explicit chiral symmetry breaking through the quark masses.
Ten additional LECs $L_i$ are needed at $\cO(p^4)$.

\begin{figure}[t]
\begin{minipage}[c]{.5\linewidth}\centering
\setlength{\unitlength}{0.46mm}          
\begin{picture}(163,141)
\put(0,0){\makebox(163,141){}}
\thicklines
\put(8,130){\makebox(25,11){Energy}}
\put(43,130){\makebox(42,11){Fields}}
\put(101,130){\makebox(52,11){Effective Theory}}
\put(5,130){\line(1,0){153}} {
\put(8,91){\makebox(25,34){$M_W$}}
\put(43,91){\framebox(42,34){$\ba W, Z, \gamma, g \\
     \tau, \mu, e, \nu_i \\ t, b, c, s, d, u \ea $}}
\put(101,91){\makebox(52,34){Standard Model}}

\put(8,45){\makebox(25,22){$\lsim m_c$}}
\put(43,45){\framebox(42,22){$\ba  \gamma, g  \; ;\; \mu ,  e, \nu_i
             \\ s, d, u \ea $}}
\put(101,45){\makebox(52,22){$\cL_{\mathrm{QCD}}^{n_f=3}$,
             $\cH_{\mathrm{eff}}^{\Delta S=1,2}$}}

\put(8,0){\makebox(25,22){$m_K$}}
\put(43,0){\framebox(42,22){$\ba\gamma \; ;\; \mu , e, \nu_i  \\
            \pi, K,\eta  \ea $}}
\put(101,0){\makebox(52,22){$\chi$PT}}
\linethickness{0.3mm}
\put(64,41){\vector(0,-1){15}}
\put(64,86){\vector(0,-1){15}}
\put(69,76){OPE}
\put(69,31){$N_C\to\infty $}}                     
\end{picture}
\vskip -.3cm\mbox{}
\caption{Evolution from $M_W$ to the kaon mass scale.
  \label{fig:eff_th}}
\end{minipage}
\hfill
\begin{minipage}[c]{.43\linewidth}\centering
\mbox{}\vskip -.5cm
\includegraphics[width=6.2cm]{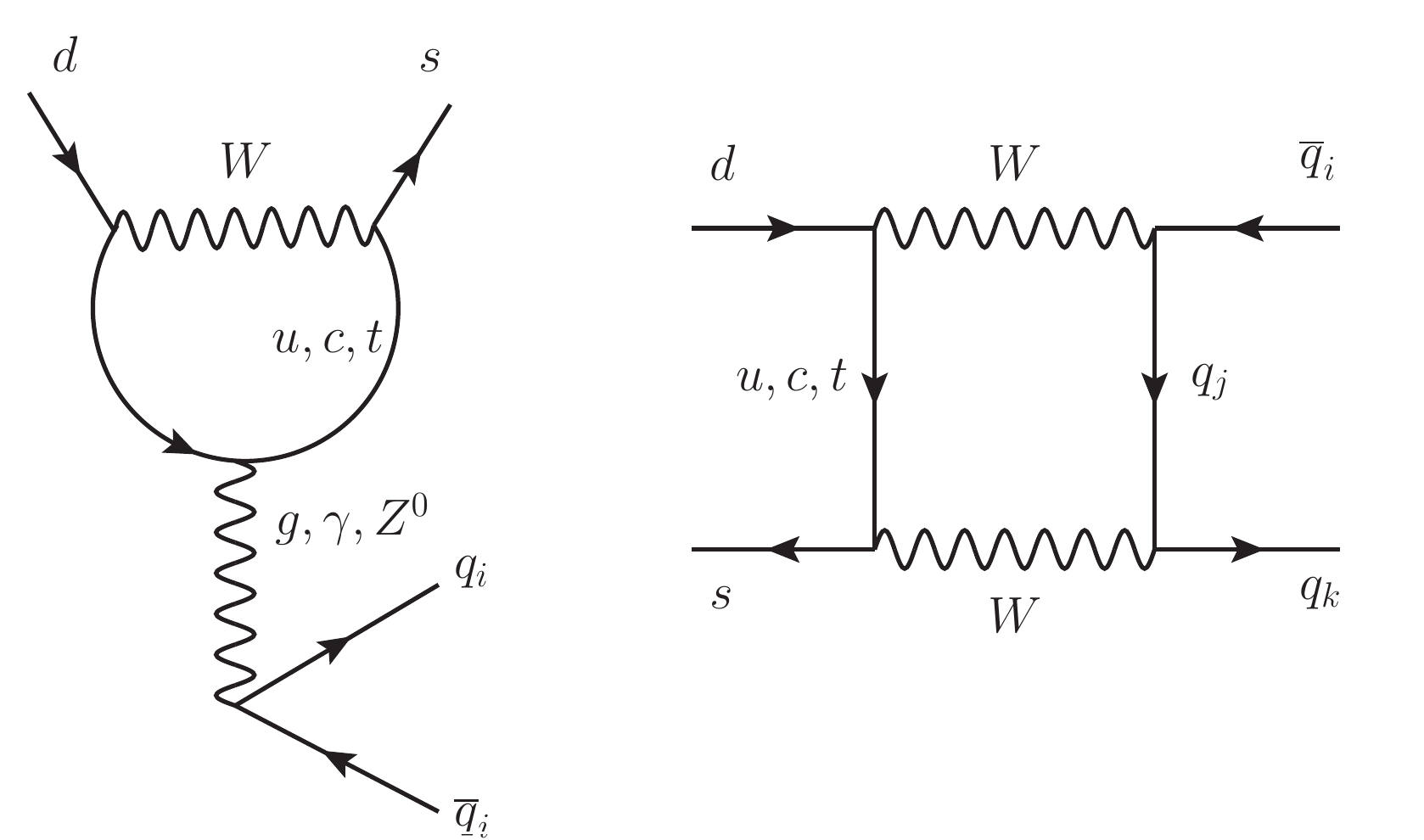}
\caption{\label{fig:sd-diagrams}
Short-distance Feynman diagrams.}
\vskip .3cm
\includegraphics[width=6.5cm]{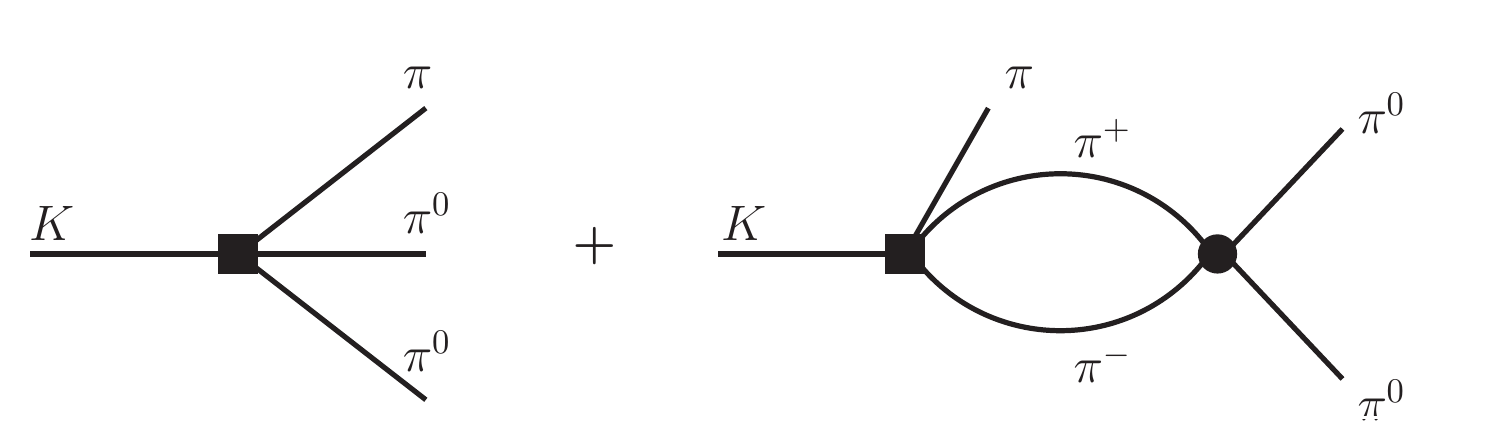}
\caption{\label{fig:sd-diagrams}
Long-distance Feynman diagrams.}
\end{minipage}
\end{figure}

Non-leptonic weak interactions with $\Delta S=1$ are incorporated as a perturbation to
the strong $\chi$PT Lagrangian. 
At LO the most general effective Lagrangian, with the same
$SU(3)_L\otimes SU(3)_R$ transformation properties as the short-distance
Hamiltonian \eqn{eq:sd_hamiltonian}, contains three terms \cite{Cirigliano:2011ny}:
\bel{eq:lg8_g27}
\cL_2^{\Delta S=1} = -{G_F \over \sqrt{2}}  V_{ud}^{\phantom{*}} V_{us}^*
\left\{ g_8  \,\langle\lambda L_{\mu} L^{\mu}\rangle   +
g_{27} \left( L_{\mu 23} L^\mu_{11} + {2\over 3} L_{\mu 21} L^\mu_{13}\right) +
e^2 g_8  g_{\rms ew} F^6\, \langle\lambda U^\dagger Q U\rangle +
\mbox{\rm h.c.} \right\} ,
\ee
where $L_{\mu}=i F^2 U^\dagger D_\mu U$  represents the octet of
$V-A$ currents, $\lambda\equiv (\lambda_6 - i \lambda_7)/2$ projects onto the
$\bar s\to \bar d$ transition, 
$Q={1\over 3} \,\mbox{\rm diag}(2,-1,-1)$ is the quark charge matrix
and $\langle\cdots \rangle$ denotes the 3-dimensional flavour trace.
The LECs $g_8$ and $g_{27}$ measure the strength of the two
parts of $\cH_{\mbox{\rms eff}}^{\Delta S=1}$  
transforming as
$(8_L,1_R)$ and $(27_L,1_R)$, respectively, under chiral rotations, while $g_{\rms ew}$
accounts for the electromagnetic penguin operators.

The $\chi$PT framework determines the most general form of the
K decay amplitudes, compatible with chiral symmetry,
in terms of the LECs multiplying the relevant chiral operators.
These LECs, which encode the short-distance dynamics,
can be determined phenomenologically and/or calculated
in the limit of a large number of QCD colours (matching).
Chiral loops generate non-polynomial contributions, with
logarithms and threshold factors as required by unitarity.
Fig.~\ref{fig:eff_th} shows schematically the procedure used
to evolve down from $M_W$ to $m_K$.
While the OPE resums the short-distance logarithmic corrections
$\log{(M/\mu)}$, the $\chi$PT loops take care of the large
infrared logarithms  $\log{(\mu/m_\pi)}$ associated with unitarity corrections
(final-state interactions).

\section{Leptonic and Semileptonic Decays.}

Leptonic and semileptonic kaon decays are well understood theoretically,
including electromagnetic corrections. Strong interactions only appear through the
hadronic matrix elements of the left-handed weak current, which can be precisely studied
within $\chi$PT and with lattice simulations.

The ratios \
$R^{(P)}_{e/\mu}\equiv \Gamma[P^-\to e^-\bar\nu_e (\gamma)]/\Gamma[P^-\to \mu^-\bar\nu_\mu (\gamma)]$ \
($P=\pi,K$) have been calculated \cite{Marciano:1993sh,Cirigliano:2007xi}
and measured \cite{Britton:1993cj,Czapek:1993kc,NA62:2011aa,Ambrosino:2009aa,Bucci} with high accuracy, allowing for a test
of charged-current lepton universality at the 0.2\% level.
As shown in Table~1, similar precisions have been achieved in $K\to\pi\ell\nu_\ell$ \cite{Antonelli:2010yf}
and $\tau\to\nu_\tau\ell\nu_\ell$ \cite{Pich:2011nh} decays.

The ratio of radiative inclusive decay rates
$\Gamma[K^-\to \mu^-\bar\nu_\mu (\gamma)]/\Gamma[\pi^-\to \mu^-\bar\nu_\mu (\gamma)]$
provides information on the quark mixing matrix \cite{Antonelli:2010yf,Marciano:2004uf}.
With a careful treatment ot electromagnetic and isospin-violating corrections,
one extracts $|V_{us}/V_{ud}| |F_K/F_\pi| = 0.2763 \pm 0.0005$ \cite{Cirigliano:2011tm}. Taking for the ratio of meson
decay constants the lattice average $F_K/F_\pi = 1.193 \pm 0.006$ \cite{Colangelo:2010et}, this gives
\be\label{eq:Vus_ud}
|V_{us}/V_{ud}| \; =\; 0.2316 \pm 0.0012\, .
\ee
%

\begin{figure}[t]\centering
\begin{minipage}[c]{.45\linewidth}\centering
\begin{tabular}{ccc}
\hline\hline
$|g_\mu/g_e|$  && Source\\ \hline
$1.0021\pm 0.0016$ &\qquad & $\pi\to \mu/e$
\\
$0.9978\pm 0.0018$ &&  $K\to \mu/e$
\\
$1.0010\pm 0.0025$ &&  $K\to \pi\,\mu/e$
\\
$1.0018\pm 0.0014$ &&  $\tau\to \mu/e $
\\ \hline\hline
\end{tabular}
\vskip .2cm
\vbox{\small\begin{flushleft} {\bf Table 1:} Determinations of the ratio
of the $\mu^-$ and $e^-$ couplings to the $W$. \end{flushleft}}
\end{minipage}
\hfill
\begin{minipage}[c]{.5\linewidth}\centering
\includegraphics[width=6.5cm]{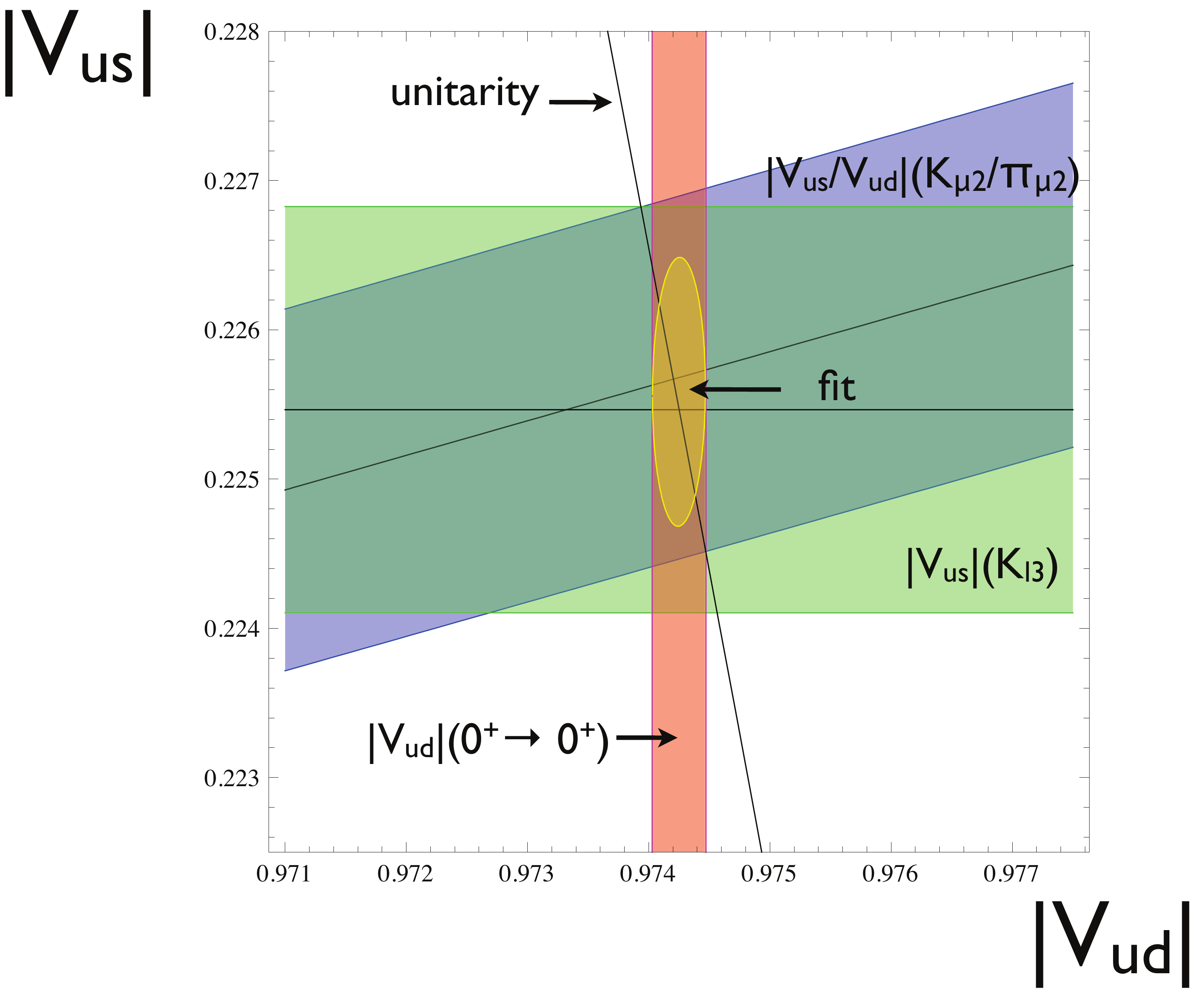}
\caption{\label{fig:unitarity}
Constraints on $|V_{ud}|$ and $|V_{us}|$ from $K_{\ell 3}$ decays (horizontal),
$0^+ \to 0^+$ nuclear decays (vertical) and $K_{\mu 2}/\pi_{\mu 2}$ (oblique band) \cite{Cirigliano:2011tm}.}
\end{minipage}
\end{figure}

The most recent 
$K_{\ell 3}$ experiments have resulted in
improved precision and significant shifts in the branching fractions \cite{PDG}.
Including electromagnetic and isospin-breaking corrections \cite{Cirigliano:2008wn,Kastner:2008ch},
one obtains
$|V_{us}\, f_+(0)| = 0.2163\pm 0.0005$ \cite{Antonelli:2010yf},
with $f_+(0)= 1 + \cO[(m_s-m_u)^2]$ the $K^0\to\pi^-\ell^+\nu_\ell$ vector form factor.
The exact value of $f_+(0)$ has been thoroughly investigated since the first
precise estimate by Leutwyler and Roos, $f_+(0) = 0.961\pm 0.008$ \cite{Leutwyler:1984je}.
While analytical calculations based on $\chi$PT obtain
higher values \cite{Jamin:2004re,Cirigliano:2005xn},
owing to the large 
($\sim 0.01$) 2-loop corrections \cite{Bijnens:2003uy},
lattice results \cite{Colangelo:2010et} tend to agree with the Leutwyler--Roos estimate.
Taking as reference value the most recent and precise lattice result
\cite{Boyle:2010bh}, $f_+(0) = 0.960\pm 0.006$, one obtains \cite{Cirigliano:2011ny}
\bel{eq:Vus}
|V_{us}|\, =\, 0.2255\pm 0.0005_{\rms exp}\pm 0.0012_{\rms th} \, .
\ee
Together with $|V_{ud}| = 0.97425 \pm 0.00022$
\cite{Towner:2010zz}
and the negligible $|V_{ub}|$ contribution \cite{PDG}, the determinations 
\eqn{eq:Vus_ud} and \eqn{eq:Vus} imply an stringent test of the unitarity of the
quark mixing matrix \cite{Cirigliano:2011tm}:
\be
\Delta_{\rms CKM} = |V_{ud}|^2  +  |V_{us}|^2  +  |V_{ub}|^2  - 1 \; =\;  0.0001 \pm 0.0006\, .
\ee

\section{Nonleptonic Decays and Direct CP Violation: $\varepsilon'/\varepsilon$}

The measured $A(K\to\pi\pi)_I$ decay amplitudes show a strong enhancement of the octet
$\Delta I =\frac{1}{2}$ transition amplitude into a $2\pi$ final state with isospin $I=0$:
$|A_0/A_2|\approx 22$. In the $\chi$PT framework
this manifests as a huge difference between the LECs in Eq.~\eqn{eq:lg8_g27}. A LO fit to
the data gives
$\vert g_8 \vert \simeq 5.0$ and $\vert g_{27} \vert \simeq 0.285$.
Part of the enhancement originates in the strong rescattering of the final pions, which at one loop increases $A_0$
by roughly 35\%; taking the $\chi$PT 1-loop contributions into account, one finds a sizeably smaller
octet coupling
(the central values change slightly to 3.61 and 0.297, respectively, if isospin violation is included)
\cite{Cirigliano:2011ny,Cirigliano:2009rr,Cirigliano:2003gt}:
\be
\vert g_8 \vert\, =\, 3.62\pm 0.28\, ,\qquad \qquad\qquad
\vert g_{27} \vert \, =\, 0.286\pm 0.28\, .
\ee
In the absence of QCD corrections, the SM ($W$ exchange) implies $g_8 = g_{27} = \frac{3}{5}$, very far
from the phenomenologically required values. The computed short-distance QCD corrections show the needed
qualitative trend to understand the data, but
a proper calculation of the hadronic matrix elements of the relevant $Q_i$ operators is still lacking.
The matching of the effective descriptions $\cH_{\mbox{\rms eff}}^{\Delta S=1}$ (short-distance) and
$\cL_2^{\Delta S=1}$ ($\chi$PT) can be done in the large-$N_C$ limit with the result
$g_8^\infty = 1.13 \pm 0.18$ and $g_{27}^\infty = 0.46 \pm 0.01$ \cite{Cirigliano:2011ny}, which shows the relevance of the missing
NLO corrections in $1/N_C$.
Lattice simulations have recently achieved a quite successful description of $A_2$ ($g_{27}$) \cite{Blum:2011ng},
but a real quantitative understanding of $A_0$ ($g_8$) remains still problematic \cite{Blum:2011pu}.

The situation is much better fot the 
ratio \cite{Batley:2002gn,Barr:1993rx,Abouzaid:2010ny,Gibbons:1993zq}
[$\eta_{_{ab}}\equiv A(K_L\to\pi^a\pi^b)/ A(K_S\to\pi^a\pi^b)$]
\be\label{eq:exp}
{\rm Re} \left(\varepsilon'/\varepsilon\right)\; =\;
\frac{1}{3} \left( 1   -\left|\frac{\eta_{_{00}}}{\eta_{_{+-}}}\right|\right) \; =\;
(16.8 \pm 2.0) \times  10^{-4} \, ,
\ee
which demonstrates the existence of direct CP violation in the $K\to 2\pi$ decay amplitudes.
When CP violation is turned on, the amplitudes $A_I$
acquire imaginary parts. To first order in CP violation,
\begin{equation}
\varepsilon'\;  = - \frac{i}{\sqrt{2}} \, e^{i ( \chi_2 - \chi_0 )} \,
\frac{\mathrm{Re} A_{2}}{ \;\mathrm{Re} A_{0}} \,
\left[
\frac{\mathrm{Im} A_{0}}{ \mathrm{Re} A_{0}} \, - \,
\frac{\mathrm{Im} A_{2}}{ \mathrm{Re} A_{2}} \right] ,
\label{eq:cp1}
\end{equation}
where the strong phases $\chi_I$ can be identified with the S-wave $\pi\pi$
scattering phase shifts at $\sqrt{s}=m_K$, up to isospin-breaking effects \cite{Cirigliano:2009rr,Cirigliano:2003gt}.
The phase $\phi_{\varepsilon'} =\chi_2 - \chi_0 + \pi/2 = (42.5\pm 0.9)^\circ$
is very close to the so-called superweak phase,
$\phi_\varepsilon \approx \tan^{-1}{
\left[
2 (m_{K_L}-m_{K_S})/(\Gamma_{K_S}-\Gamma_{K_L})
\right]}
= (43.51\pm 0.05)^\circ$,
implying that $\cos{(\phi_{\varepsilon'} -\phi_\varepsilon)}\approx 1$.
The CP-conserving amplitudes $\mathrm{Re} A_{I}$ can be set to their experimentally determined values,
avoiding in this way the large uncertainties associated with the
hadronic matrix elements of the four-quark operators in $\cH_{\mathrm{eff}}^{\Delta S=1}$.
Thus, one only needs a first-principle calculation of the CP-odd amplitudes $\mathrm{Im} A_{0}$
and $\mathrm{Im} A_{2}$; the first one is completely dominated by the strong penguin operator
$Q_6$, while the leading contribution to the second one comes from the
electromagnetic penguin $Q_8$.
Fortunately, those are precisely the only operators that are
well approximated through a large-$N_C$ estimate of LECs, because their anomalous dimensions
are leading in $1/N_C$.
Owing to the large ratio  $\mathrm{Re} A_{0}/ \mathrm{Re} A_{2}$, isospin violation
plays also an important role in $\varepsilon'/\varepsilon$ \cite{Cirigliano:2003gt}.
The one-loop $\chi$PT enhancement of the isoscalar amplitude \cite{Pallante:1999qf,Pallante:2001he} 
destroys an accidental LO cancellation of the two terms in \eqn{eq:cp1} \cite{Buchalla:1995vs,Bertolini:1998vd,Hambye:1999yy},
bringing the SM prediction
of $\varepsilon'/\varepsilon$ in good agreement with the experimental measurement in Eq.~(\ref{eq:exp})
\cite{Pallante:1999qf,Pallante:2001he,Pich:2004ee}:
\be\label{eq:finalRes}
\mbox{Re}\left(\epsilon'/\epsilon\right) \; =\;
\left(19\pm 2_{\mu}\, {{}_{-6}^{+9}}_{m_s} \pm 6_{1/N_C}\right) \times 10^{-4}\, .
\ee
%

\section{Rare and Radiative Decays}

Kaon decays mediated by flavour-changing neutral currents are suppressed in the SM and
their main interest, other than their own understanding, relies on the
possible observation of new physics effects.  Most of these processes are dominated by
long-distance contributions; however, there are also
processes governed by short-distance amplitudes, such as $K \rightarrow\pi \nu \bar{\nu}$.

\subsection{$K^0\to\gamma\gamma$ \ and \ $K^0\to\ell^+\ell^-$}

\begin{figure}[t]
\begin{minipage}[c]{.46\linewidth}\centering
\includegraphics[width=6.2cm]{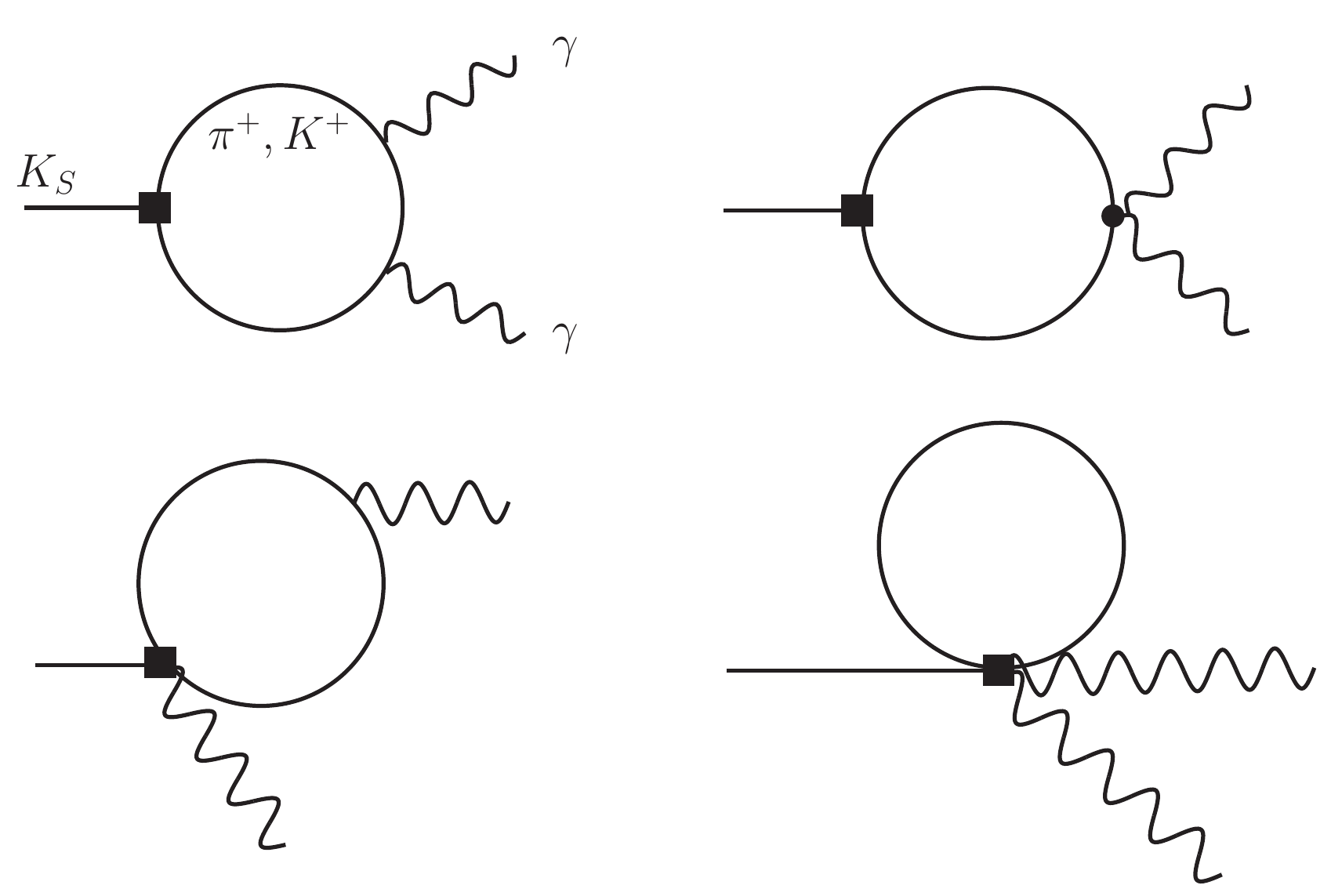}
\caption{Lowest-order 
contributions to $K_S\to\gamma\gamma$.
  \label{fig:KSgg}}
\end{minipage}
\hfill
\begin{minipage}[c]{.46\linewidth}\centering
\mbox{}\vskip -.5cm
\includegraphics[width=6.2cm]{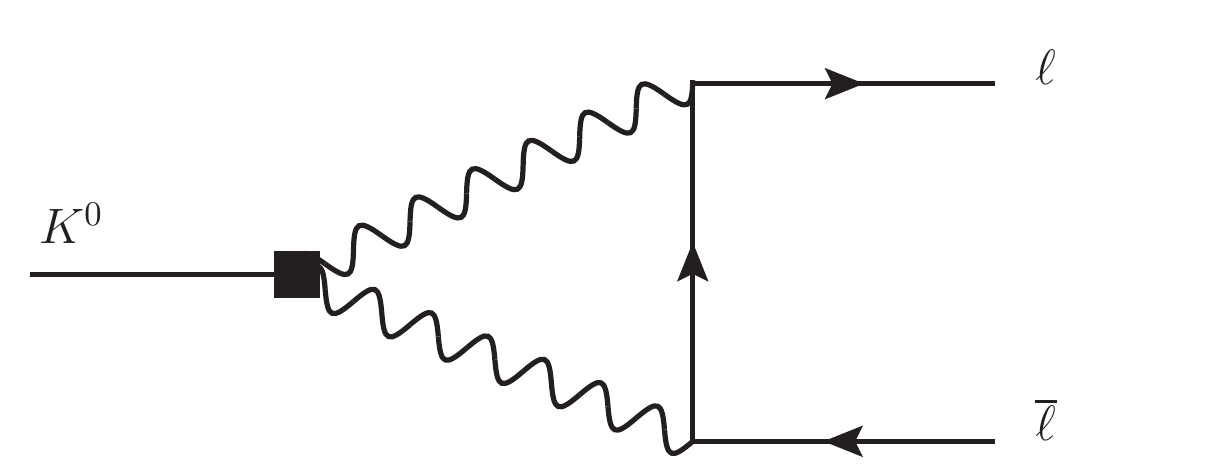}
\caption{\label{fig:KSll}
$2\gamma$ contribution to $K^0\to\ell^+\ell^-$.}
\end{minipage}
\end{figure}

The symmetry constraints do not allow any tree-level $K_1^0\gamma\gamma$ coupling at $\cO(p^4)$
($K^0_{1,2}$ are the CP-even and CP-odd states).
The  decay $K_S\to\gamma\gamma$ proceeds then through a one-loop amplitude, with intermediate $\pi^+\pi^-$,
which is necessarily finite because there are no counterterms to renormalize divergences.
The resulting $\cO(p^4)$ prediction \cite{D'Ambrosio:1986ze,Goity:1986sr},
$\mbox{\rm Br}(K_S\to\gamma\gamma) = 2.0 \times 10^{-6}$,
is slightly lower than the experimental measurement
$\mbox{\rm Br}(K_S\to\gamma\gamma)  = (2.63 \pm 0.17) \times 10^{-6}$ \cite{PDG}.
Full agreement is obtained at
$\cO(p^6)$, once  rescattering corrections ($K_S\to\pi\pi\to\pi^+\pi^-\to\gamma\gamma$) are included
 \cite{Kambor:1993tv}.


The 2-loop amplitude $K_S \rightarrow \gamma^* \gamma^* \rightarrow \ell^+
\ell^-$ is also finite  \cite{Ecker:1991ru}
because chiral symmetry forbids any CP-invariant local contribution at this order.
The predicted rates,
$\mathrm{Br}(K_S \rightarrow e^+ e^-) = 2.1\times 10^{-14}$
and
$\mathrm{Br}(K_S \rightarrow \mu^+ \mu^-) = 5.1\times 10^{-12}$  \cite{Ecker:1991ru},
are well below the experimental upper bounds
$\mathrm{Br}(K_S \rightarrow e^+ e^-) < 9\times 10^{-9}$   
and
$\mathrm{Br}(K_S \rightarrow\mu^+ \mu^-) < 3.2\times 10^{-7}$   
(90\%  C.L.) \cite{PDG}.
This calculation allows us to compute
the longitudinal polarization $P_L$ of either muon in the decay
$K_L \rightarrow \mu^+ \mu^-$, a CP-violating observable which
in the SM is dominated by indirect CP violation from
$K^0$--$\bar K^0$ mixing. One finds
$|P_L| = (2.6\pm 0.4)\times 10^{-3}$ \cite{Ecker:1991ru}.

\subsection{$K\to\pi\gamma\gamma$}

\begin{figure}[t]
\begin{minipage}[c]{.46\linewidth}\centering
\includegraphics[width=6.2cm]{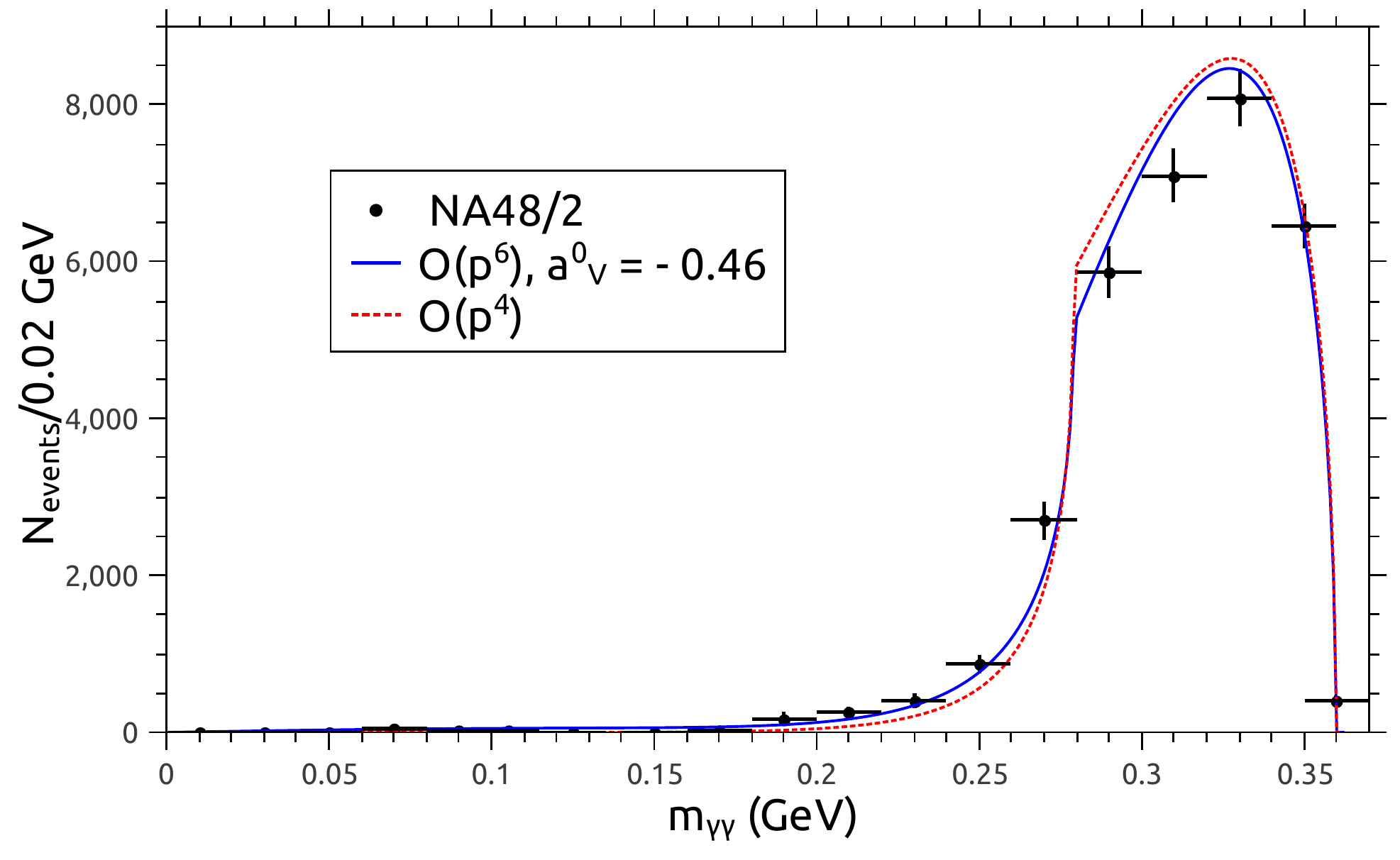}
\caption{$K_L\to\pi^0\gamma\gamma$ spectra
at $\cO(p^4)$ and $\cO(p^6)$ in
$\chi$PT. The data are from Ref.~\cite{Lai:2002kf}.
  \label{fig:KL_pgg}}
\end{minipage}
\hfill
\begin{minipage}[c]{.46\linewidth}\centering
\mbox{}\vskip -.2cm
\includegraphics[width=6.2cm]{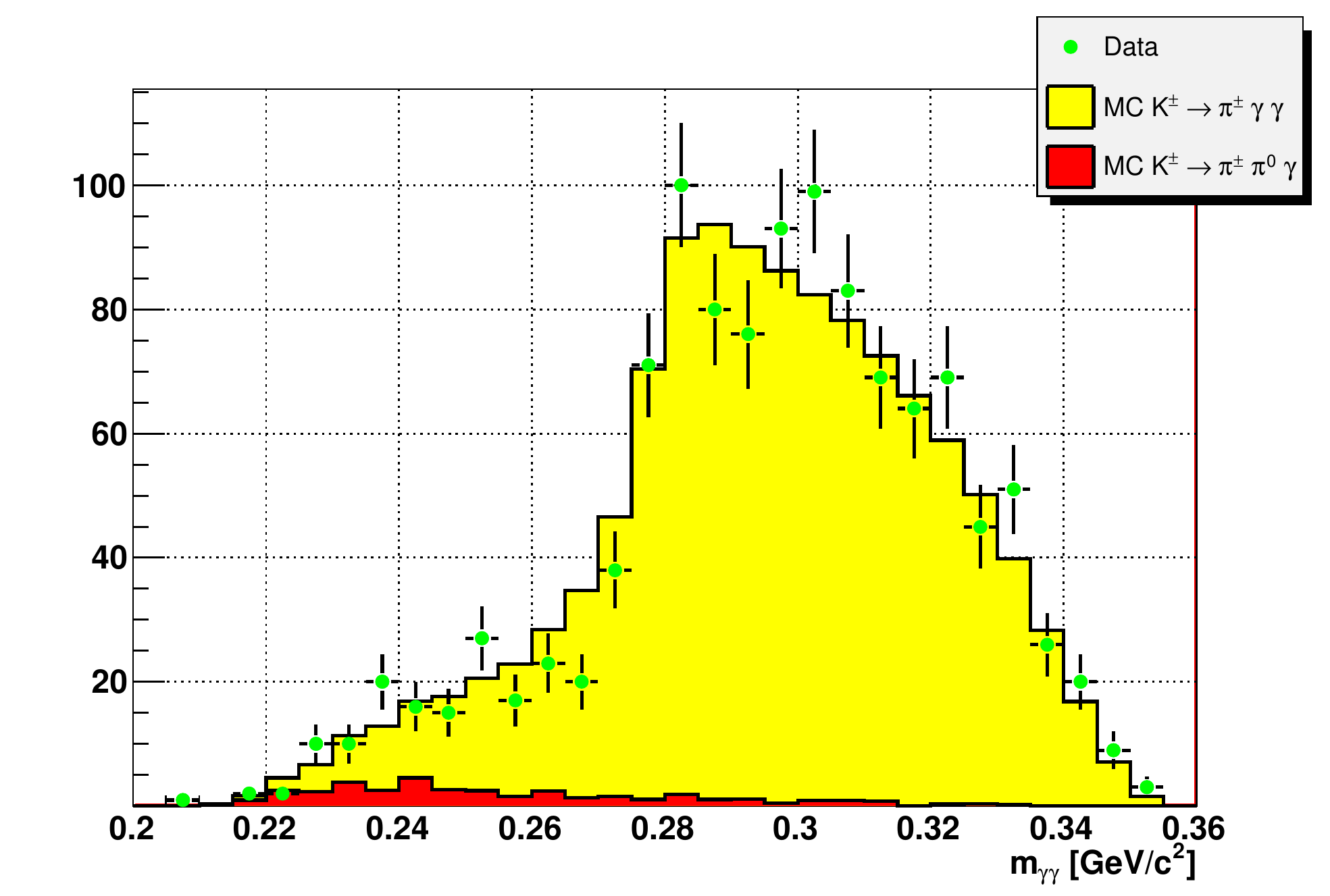}
\caption{\label{fig:K+p+gg}
Measured $K^\pm\to\pi^\pm\gamma\gamma$ spectrum \cite{Morales}.}
\end{minipage}
\end{figure}

Again, the symmetry constraints do not allow any tree-level contribution to $K_2\to\pi^0\gamma\gamma$
from $\cO(p^4)$ terms in the Lagrangian. The decay amplitude is therefore determined by a finite
loop calculation \cite{Ecker:1987fm,Cappiello:1988yg,Sehgal:1989pw}.
Due to the large absorptive $\pi^+\pi^-$ contribution, the spectrum in
the invariant mass of the two photons is predicted to have a very characteristic behaviour
(dotted line in Fig.~\ref{fig:KL_pgg}), peaked at high values of $m_{\gamma\gamma}$.
The agreement with the measured distribution \cite{Lai:2002kf} is remarkably good. However, the $\cO(p^4)$ prediction for
the rate,  $\mathrm{BR}(K_L \rightarrow \pi^0 \gamma \gamma) = 6.8 \times 10^{-7}$ \cite{Ecker:1987fm}, is
significantly smaller than the present PDG average \cite{PDG}
\be
{\rm BR}(K_L \rightarrow \pi^0 \gamma \gamma)\, =\, (1.27 \pm 0.03) \times 10^{-6}\, ,
\ee
indicating that higher-order corrections are sizeable.
Unitarity corrections from $K_L \rightarrow \pi^+ \pi^- \pi^0$
\cite{Cohen:1993ta,Cappiello:1992kk} and local vector-exchange contributions \cite{Cohen:1993ta,Ecker:1990in}
restore the agreement at ${\cal O}(p^6)$.

A quite similar spectrum is predicted \cite{Ecker:1987hd} for the charged mode $K^\pm\to\pi^\pm\gamma\gamma$,
but in this case there is a free LEC already at $\cO(p^4)$.
$\cO(p^6)$ corrections have been also investigated \cite{D'Ambrosio:1996zx}.
Both the spectrum and the rate can be correctly reproduced \cite{Morales,Fantechi}.

\subsection{$K_L\to\pi^0 e^+e^-$}

This decay 
is an interesting process in looking for new CP-violating signatures, because
$K_2^0\to\pi^0\gamma^*$ violates CP \cite{Ecker:1987hd,Donoghue:1994yt}.
The CP-conserving amplitude proceeds through a $2\gamma$ intermediate state and
is suppressed by an additional power of $\alpha$. Using the $K_L \rightarrow \pi^0 \gamma\gamma$ data,
the CP-conserving rate is found to be below $10^{-12}$ \cite{Cirigliano:2011ny}.
The $K_L \rightarrow \pi^0 e^+ e^-$ transition is then dominated by the $\cO(\alpha)$ CP-violating contributions \cite{Ecker:1987hd},
both from $K^0$--$\bar K^0$ mixing and direct CP violation. The estimated rate
$\mathrm{Br}(K_L \rightarrow \pi^0 e^+ e^-) =  (3.1  \pm 0.9)\times 10^{-11}$ \cite{Cirigliano:2011ny,Buras:1994qa,Buchalla:2003sj}
is only a factor 10 smaller than the present (90\% C.L.) upper bound of $2.8 \times 10^{-10}$ \cite{AlaviHarati:2003mr}
and should be reachable in the near future.

\subsection{$K\to\pi\nu\bar\nu$}

Long-distance effects play a negligible role in $K^+\to\pi^+\nu\bar\nu$
and $K_L\to\pi^0\nu\bar\nu$.
These processes are dominated by short-distance loops ($Z$ penguin,
$W$ box) involving the heavy top quark; the charged mode receives also sizeable
contributions from internal charm-quark exchanges.
The decay amplitudes are proportional to the hadronic matrix element of the $\Delta S=1$ vector current,
which (assuming isospin symmetry) can be obtained from $K_{\ell 3}$ decays:
\bel{eq:pnn}
T(K\to\pi\nu\bar\nu)\,\sim\, \sum_{i=c,t}
F(V_{id}^{\phantom{*}} V_{is}^*;x_i)\;
\left(\bar\nu_L\gamma_\mu\nu_L\right)\;
\langle\pi |\bar s_L\gamma^\mu d_L|K\rangle
\, ,
\qquad\qquad x_i\equiv m_i^2/M_W^2 \, .
\ee
The small long-distance and isospin-violating corrections can be estimated within $\chi$PT.
The $K_L\to\pi^0\nu\bar\nu$ transition violates CP and is completely dominated by direct CP violation, the
contribution from $K^0$--$\bar K^0$ mixing being only of the order of 1\%.
Taking the CKM inputs from global fits, one predicts
$\mathrm{Br} (K_L \to \pi^0 \nu \bar{\nu}) =  (2.4 \pm 0.4) \times 10^{-11}$
and
$\mathrm{Br} (K^+  \to \pi^+ \nu \bar{\nu}) = (0.78 \pm 0.08) \times 10^{-10}$
\cite{Buras:2005gr,Brod:2010hi}.
The uncertainties are largely parametrical,
due to CKM input, $m_{c}$, $m_{t}$  and $\alpha_{s} (M_Z)$.

The $K^+$ mode has been observed \cite{Artamonov:2008qb}, while only an upper bound exists on the $K_L$ one \cite{Ahn:2009gb}:
\be
\mathrm{Br} (K^+  \to \pi^+ \nu \bar{\nu})
 =  (1.73^{+1.15}_{-1.05}) \times 10^{-10}\, ,
\qquad\;
\mathrm{Br} (K_L \to \pi^0 \nu \bar{\nu})
 <   2.6  \times 10^{-8}  \;\:  (90 \% \, {\rm C.L.})\, .
\ee
New experiments, aiming to reach $\mathcal{O}(100)$ events (assuming SM rates),
are under development at CERN (NA62) and J-PARC (K0TO)
for charged and neutral modes, respectively.
Increased sensitivities could be obtained through the recent ORKA proposal for a
$K^+  \to \pi^+ \nu \bar{\nu}$ experiment at Fermilab and the higher kaon fluxes available at Project-X
\cite{FNAL}.

\acknowledgments

Work supported in part by the Spanish Government~[grants FPA2007-60323, FPA2011-23778 and CSD2007-00042~(Consolider Project CPAN)], the Generalitat Valenciana [Prometeo/2008/069]
and the Alexander von Humboldt Foundation.


\begin{thebibliography}{99}

\bibitem{Cirigliano:2011ny}
  V.~Cirigliano, G.~Ecker, H.~Neufeld, A.~Pich and J.~Portol\'es,
  Rev.\ Mod.\ Phys.\  {\bf 84} (2012) 399
  [arXiv:1107.6001 [hep-ph]].

\bibitem{GellMann:1953zza}
  M.~Gell-Mann,
  Phys.\ Rev.\  {\bf 92} (1953) 833.

\bibitem{Pais:1952zz}
  A.~Pais,
  Phys.\ Rev.\  {\bf 86} (1952) 663.

\bibitem{Dalitz:1954cq}
  R.~H.~Dalitz,
  Phys.\ Rev.\  {\bf 94} (1954) 1046.

\bibitem{Lee:1956qn}
  T.~D.~Lee and C.~-N.~Yang,
  Phys.\ Rev.\  {\bf 104} (1956) 254.

\bibitem{Lande:1956pf} K.~Lande {\it et al.},
  Phys.\ Rev.\  {\bf 103} (1956) 1901.

\bibitem{Fry:1956pg}
  W.~F.~Fry, J.~Schneps and M.~S.~Swami,
  Phys.\ Rev.\  {\bf 103} (1956) 1904.

\bibitem{Cabibbo:1963yz}
  N.~Cabibbo,
  Phys.\ Rev.\ Lett.\  {\bf 10} (1963) 531.

\bibitem{Kobayashi:1973fv}
  M.~Kobayashi and T.~Maskawa,
  Prog.\ Theor.\ Phys.\  {\bf 49} (1973) 652.

\bibitem{Christenson:1964fg}
  J.~H.~Christenson, J.~W.~Cronin, V.~L.~Fitch and R.~Turlay,
  Phys.\ Rev.\ Lett.\  {\bf 13} (1964) 138.

\bibitem{Glashow:1970gm}
  S.~L.~Glashow, J.~Iliopoulos and L.~Maiani,
  Phys.\ Rev.\ D {\bf 2} (1970) 1285.


\bibitem{Ambrose:1998us}
  D.~Ambrose {\it et al.}  [BNL Collaboration],
  Phys.\ Rev.\ Lett.\  {\bf 81} (1998) 5734
  [hep-ex/9811038].

\bibitem{AlaviHarati:2002eh}
  A.~Alavi-Harati {\it et al.}  [KTeV Collaboration],
  Phys.\ Rev.\ Lett.\  {\bf 90} (2003) 141801
  [hep-ex/0212002].

\bibitem{Sher:2005sp}  A.~Sher {\it et al.},
  Phys.\ Rev.\ D {\bf 72} (2005) 012005
  [hep-ex/0502020].

\bibitem{Appel:2000tc}  R.~Appel {\it et al.},
  Phys.\ Rev.\ Lett.\  {\bf 85} (2000) 2877
  [hep-ex/0006003].

\bibitem{Gilman:1979bc}
  F.~J.~Gilman and M.~B.~Wise,
  Phys.\ Rev.\ D {\bf 20} (1979) 2392,
%
{\bf 21} (1980) 3150.

\bibitem{Buras:1992tc}  A.~J.~Buras {\it et al.},
  Nucl.\ Phys.\ B {\bf 400} (1993) 37
  [hep-ph/9211304],
%
75
  [hep-ph/9211321],
%
{\bf 408} (1993) 209
  [hep-ph/9303284];
%
  Phys.\ Lett.\ B {\bf 389} (1996) 749
  [hep-ph/9608365].

\bibitem{Ciuchini:1995cd} M.~Ciuchini {\it et al.},
  Z.\ Phys.\ C {\bf 68} (1995) 239
  [hep-ph/9501265];
%
  Phys.\ Lett.\ B {\bf 301} (1993) 263
  [hep-ph/9212203];
%
  Nucl.\ Phys.\ B {\bf 415} (1994) 403
  [hep-ph/9304257].


\bibitem{Weinberg:1978kz}
  S.~Weinberg,
  Physica A {\bf 96} (1979) 327.

\bibitem{Gasser:1984gg}
  J.~Gasser and H.~Leutwyler,
  Nucl.\ Phys.\ B {\bf 250} (1985) 465,
%
517;
%
  Annals Phys.\  {\bf 158} (1984) 142.

\bibitem{Ecker:1994gg}
  G.~Ecker,
  Prog.\ Part.\ Nucl.\ Phys.\  {\bf 35} (1995) 1
  [hep-ph/9501357].

\bibitem{Pich:1995bw}
  A.~Pich,
  Rept.\ Prog.\ Phys.\  {\bf 58} (1995) 563
  [hep-ph/9502366].

\bibitem{Marciano:1993sh}
  W.~J.~Marciano and A.~Sirlin,
  Phys.\ Rev.\ Lett.\  {\bf 71} (1993) 3629.

\bibitem{Cirigliano:2007xi}
  V.~Cirigliano and I.~Rosell,
  Phys.\ Rev.\ Lett.\  {\bf 99} (2007) 231801
  [arXiv:0707.3439 [hep-ph]];
%
  JHEP {\bf 0710} (2007) 005
  [arXiv:0707.4464 [hep-ph]].

\bibitem{Britton:1993cj}  D.~I.~Britton {\it et al.},
  Phys.\ Rev.\ D {\bf 49} (1994) 28;
%
  Phys.\ Rev.\ Lett.\  {\bf 68} (1992) 3000.

\bibitem{Czapek:1993kc} G.~Czapek {\it et al.},
  Phys.\ Rev.\ Lett.\  {\bf 70} (1993) 17.

\bibitem{NA62:2011aa}
  C.~Lazzeroni {\it et al.}  [NA62 Collaboration],
  Phys.\ Lett.\ B {\bf 698} (2011) 105
  [arXiv:1101.4805 [hep-ex]].

\bibitem{Ambrosino:2009aa}
  F.~Ambrosino {\it et al.}  [KLOE Collaboration],
  Eur.\ Phys.\ J.\ C {\bf 64} (2009) 627
   [Erratum-ibid.\  {\bf 65} (2010) 703]
  [arXiv:0907.3594 [hep-ex]].

\bibitem{Bucci} F. Bucci, these proceedings.

\bibitem{Antonelli:2010yf} M.~Antonelli {\it et al.},
  Eur.\ Phys.\ J.\ C {\bf 69} (2010) 399
  [arXiv:1005.2323 [hep-ph]].

\bibitem{Pich:2011nh}
  A.~Pich,
  arXiv:1112.4094 [hep-ph].

\bibitem{Marciano:2004uf}
  W.~J.~Marciano,
  Phys.\ Rev.\ Lett.\  {\bf 93} (2004) 231803
  [hep-ph/0402299].

\bibitem{Cirigliano:2011tm}
  V.~Cirigliano and H.~Neufeld,
  Phys.\ Lett.\ B {\bf 700} (2011) 7
  [arXiv:1102.0563 [hep-ph]].

\bibitem{Colangelo:2010et} G.~Colangelo {\it et al.},
  Eur.\ Phys.\ J.\ C {\bf 71} (2011) 1695
  [arXiv:1011.4408 [hep-lat]].

\bibitem{Towner:2010zz}
  I.~S.~Towner and J.~C.~Hardy,
  Rept.\ Prog.\ Phys.\  {\bf 73} (2010) 046301.

\bibitem{PDG}
J. Beringer {\it et al.} (Particle Data Group), Phys. Rev. {\bf D86} (2012) 010001.

\bibitem{Cirigliano:2008wn}
  V.~Cirigliano, M.~Giannotti and H.~Neufeld,
  JHEP {\bf 0811} (2008) 006
  [arXiv:0807.4507 [hep-ph]].

\bibitem{Kastner:2008ch}
  A.~Kastner and H.~Neufeld,
  Eur.\ Phys.\ J.\ C {\bf 57} (2008) 541
  [arXiv:0805.2222 [hep-ph]].

\bibitem{Leutwyler:1984je}
  H.~Leutwyler and M.~Roos,
  Z.\ Phys.\ C {\bf 25} (1984) 91.

\bibitem{Jamin:2004re}
  M.~Jamin, J.~A.~Oller and A.~Pich,
  JHEP {\bf 0402} (2004) 047
  [hep-ph/0401080].

\bibitem{Cirigliano:2005xn}
  V.~Cirigliano, G.~Ecker, M.~Eidemuller, R.~Kaiser, A.~Pich and J.~Portol\'es,
  JHEP {\bf 0504} (2005) 006
  [hep-ph/0503108].

\bibitem{Bijnens:2003uy}
  J.~Bijnens and P.~Talavera,
  Nucl.\ Phys.\ B {\bf 669} (2003) 341
  [hep-ph/0303103].

\bibitem{Boyle:2010bh}
  P.~A.~Boyle {\it et al.}  [RBC-UKQCD Collaboration],
  Eur.\ Phys.\ J.\ C {\bf 69} (2010) 159
  [arXiv:1004.0886 [hep-lat]].

\bibitem{Cirigliano:2009rr}
  V.~Cirigliano, G.~Ecker and A.~Pich,
  Phys.\ Lett.\ B {\bf 679} (2009) 445
  [arXiv:0907.1451 [hep-ph]].

\bibitem{Cirigliano:2003gt}
  V.~Cirigliano, G.~Ecker, H.~Neufeld and A.~Pich,
  Eur.\ Phys.\ J.\ C {\bf 33} (2004) 369
  [hep-ph/0310351];
  Phys.\ Rev.\ Lett.\  {\bf 91} (2003) 162001
  [hep-ph/0307030].


\bibitem{Blum:2011ng} T.~Blum {\it et al.},
  Phys.\ Rev.\ Lett.\  {\bf 108} (2012) 141601
  [arXiv:1111.1699 [hep-lat]];
%
  arXiv:1206.5142 [hep-lat].

\bibitem{Blum:2011pu} T.~Blum {\it et al.},
  Phys.\ Rev.\ D {\bf 84} (2011) 114503
  [arXiv:1106.2714 [hep-lat]].


\bibitem{Batley:2002gn}
  J.~R.~Batley {\it et al.}  [NA48 Collaboration],
  Phys.\ Lett.\ B {\bf 544} (2002) 97
  [hep-ex/0208009],
%
 {\bf 465} (1999) 335
  [hep-ex/9909022];
%
  Eur.\ Phys.\ J.\ C {\bf 22} (2001) 231
  [hep-ex/0110019].

\bibitem{Barr:1993rx}
  G.~D.~Barr {\it et al.}  [NA31 Collaboration],
  Phys.\ Lett.\ B {\bf 317} (1993) 233,
%
{\bf 206} (1988) 169.


\bibitem{Abouzaid:2010ny}
  E.~Abouzaid {\it et al.}  [KTeV Collaboration],
  Phys.\ Rev.\ D {\bf 83} (2011) 092001
  [arXiv:1011.0127 [hep-ex]],
%
{\bf 67} (2003) 012005
   [Erratum-ibid.\ D {\bf 70} (2004) 079904]
  [hep-ex/0208007];
%
  Phys.\ Rev.\ Lett.\  {\bf 83} (1999) 22
  [hep-ex/9905060].

\bibitem{Gibbons:1993zq} L.~K.~Gibbons {\it et al.},
  Phys.\ Rev.\ Lett.\  {\bf 70} (1993) 1203.

\bibitem{Pallante:1999qf}
  E.~Pallante and A.~Pich,
  Phys.\ Rev.\ Lett.\  {\bf 84} (2000) 2568
  [hep-ph/9911233];
%
  Nucl.\ Phys.\ B {\bf 592} (2001) 294
  [hep-ph/0007208].

\bibitem{Pallante:2001he}
  E.~Pallante, A.~Pich and I.~Scimemi,
  Nucl.\ Phys.\ B {\bf 617} (2001) 441
  [hep-ph/0105011].

\bibitem{Buchalla:1995vs}
  G.~Buchalla, A.~J.~Buras and M.~E.~Lautenbacher,
  Rev.\ Mod.\ Phys.\  {\bf 68} (1996) 1125
  [hep-ph/9512380].

\bibitem{Bertolini:1998vd}
  S.~Bertolini, M.~Fabbrichesi and J.~O.~Eeg,
  Rev.\ Mod.\ Phys.\  {\bf 72} (2000) 65
  [hep-ph/9802405].

\bibitem{Hambye:1999yy} T.~Hambye {\it et al.},
  Nucl.\ Phys.\ B {\bf 564} (2000) 391
  [hep-ph/9906434].

\bibitem{Pich:2004ee}
  A.~Pich,
  hep-ph/0410215.


\bibitem{D'Ambrosio:1986ze}
  G.~D'Ambrosio and D.~Espriu,
  Phys.\ Lett.\ B {\bf 175} (1986) 237.

\bibitem{Goity:1986sr}
  J.~L.~Goity,
  Z.\ Phys.\ C {\bf 34} (1987) 341.

\bibitem{Kambor:1993tv}
  J.~Kambor and B.~R.~Holstein,
  Phys.\ Rev.\ D {\bf 49} (1994) 2346
  [hep-ph/9310324].

\bibitem{Ecker:1991ru}
  G.~Ecker and A.~Pich,
  Nucl.\ Phys.\ B {\bf 366} (1991) 189.

\bibitem{Ecker:1987fm}
  G.~Ecker, A.~Pich and E.~de Rafael,
  Phys.\ Lett.\ B {\bf 189} (1987) 363.

\bibitem{Cappiello:1988yg}
  L.~Cappiello and G.~D'Ambrosio,
  Nuovo Cim.\ A {\bf 99} (1988) 155.

\bibitem{Sehgal:1989pw}
  L.~M.~Sehgal,
  Phys.\ Rev.\ D {\bf 41} (1990) 161.

\bibitem{Lai:2002kf}
  A.~Lai {\it et al.}  [NA48 Collaboration],
  Phys.\ Lett.\ B {\bf 536} (2002) 229
  [hep-ex/0205010].

\bibitem{Cohen:1993ta}
  A.~G.~Cohen, G.~Ecker and A.~Pich,
  Phys.\ Lett.\ B {\bf 304} (1993) 347.

\bibitem{Cappiello:1992kk}
  L.~Cappiello, G.~D'Ambrosio and M.~Miragliuolo,
  Phys.\ Lett.\ B {\bf 298} (1993) 423.

\bibitem{Ecker:1990in}
  G.~Ecker, A.~Pich and E.~de Rafael,
  Phys.\ Lett.\ B {\bf 237} (1990) 481.

\bibitem{Ecker:1987hd}
  G.~Ecker, A.~Pich and E.~de Rafael,
  Nucl.\ Phys.\ B {\bf 303} (1988) 665.


\bibitem{D'Ambrosio:1996zx}
  G.~D'Ambrosio and J.~Portol\'es,
  Phys.\ Lett.\ B {\bf 386} (1996) 403
  [hep-ph/9606213];
%
  Nucl.\ Phys.\ B {\bf 492} (1997) 417
  [hep-ph/9610244].

\bibitem{Morales}
C. Morales (NA48/2), arXiv:0805.3312 [hep-ex].

\bibitem{Fantechi} R. Fantechi, these proceedings.

\bibitem{Donoghue:1994yt}
  J.~F.~Donoghue and F.~Gabbiani,
  Phys.\ Rev.\ D {\bf 51} (1995) 2187
  [hep-ph/9408390].

\bibitem{Buras:1994qa}
  A.~J.~Buras, M.~E.~Lautenbacher, M.~Misiak and M.~Munz,
  Nucl.\ Phys.\ B {\bf 423} (1994) 349
  [hep-ph/9402347].

\bibitem{Buchalla:2003sj}
  G.~Buchalla, G.~D'Ambrosio and G.~Isidori,
  Nucl.\ Phys.\ B {\bf 672} (2003) 387
  [hep-ph/0308008].

\bibitem{AlaviHarati:2003mr}
  A.~Alavi-Harati {\it et al.}  [KTeV Collaboration],
  Phys.\ Rev.\ Lett.\  {\bf 93} (2004) 021805
  [hep-ex/0309072].

\bibitem{Buras:2005gr}
  A.~J.~Buras, M.~Gorbahn, U.~Haisch and U.~Nierste,
  Phys.\ Rev.\ Lett.\  {\bf 95} (2005) 261805
  [hep-ph/0508165];
%
  JHEP {\bf 0611} (2006) 002
  [hep-ph/0603079].

\bibitem{Brod:2010hi}
  J.~Brod, M.~Gorbahn and E.~Stamou,
  Phys.\ Rev.\ D {\bf 83} (2011) 034030
  [arXiv:1009.0947 [hep-ph]].

\bibitem{Artamonov:2008qb}
  A.~V.~Artamonov {\it et al.}  [E949 Collaboration],
  Phys.\ Rev.\ Lett.\  {\bf 101} (2008) 191802
  [arXiv:0808.2459 [hep-ex]].

\bibitem{Ahn:2009gb}
  J.~K.~Ahn {\it et al.}  [E391a Collaboration],
  Phys.\ Rev.\ D {\bf 81} (2010) 072004
  [arXiv:0911.4789 [hep-ex]];
%
  Phys.\ Rev.\ Lett.\  {\bf 100} (2008) 201802
  [arXiv:0712.4164 [hep-ex]].

\bibitem{FNAL} M. Fiorini, J. Comfort and A. Norman, these proceedings.


\end{thebibliography}
\end{document}